\documentclass[sigconf]{acmart}

\AtBeginDocument{%
  }

\setcopyright{acmlicensed}
\copyrightyear{2018}
\acmYear{2018}
\acmDOI{XXXXXXX.XXXXXXX}
\acmConference[RecSys '20]{Make sure to enter the correct
  conference title from your rights confirmation email}{September 28 - October 02,
  2026}{Minneapolis, MN}
\acmISBN{978-1-4503-XXXX-X/2018/06}

\usepackage{comment}
\usepackage{booktabs}
\usepackage{amsmath}
\usepackage{flushend}

\begin{document}

\title[]{Bridging the Semantic-Collaborative Gap: An Asymmetric Graph Architecture for Cold-Start Item Recommendation}

\author{Anh Truong}
\affiliation{%
  \institution{Tubi}
  \city{San Francisco}
  \country{USA}}
 \email{atruong@tubi.tv}

\author{John Trenkle}
\affiliation{%
  \institution{Tubi}
  \city{San Francisco}
  \country{USA}}
 \email{jtrenkle@tubi.tv}

\author{Yuanbo Chen}
\affiliation{%
  \institution{Tubi}
  \city{San Francisco}
  \country{USA}}
 \email{yuanbochen@tubitv.com}

\author{Honghong Zhao}
\affiliation{%
  \institution{Tubi}
  \city{San Francisco}
  \country{USA}}
 \email{honghongzhao@tubitv.com}

\author{Abdullah Alchihabi}
\affiliation{%
  \institution{Kumo AI}
  \city{San Francisco}
  \country{USA}}
 \email{abdullah.alchihabi@kumo.ai}

\author{Effy Fang}
\affiliation{%
  \institution{Kumo AI}
  \city{San Francisco}
  \country{USA}}
 \email{effy@kumo.ai}

\author{Michael Tamir}
\affiliation{%
  \institution{Tubi}
  \city{San Francisco}
  \country{USA}}
 \email{mtamir@tubi.tv}

\renewcommand{\shortauthors}{Truong et al.}

\begin{abstract}

Collaborative filtering and graph-based recommendation models are highly effective because they leverage observed user interactions, but this dependence creates a fundamental cold-start challenge when newly added content has no interaction history. In Tubi’s production retrieval system, this challenge is further constrained by the serving interface: new content must be assigned a standalone embedding immediately, and the model must also produce device embeddings suitable for approximate nearest-neighbor retrieval. We address this setting by formulating cold-start recommendation as an inductive graph-completion problem on a temporal bipartite device–content graph. We propose Shallow-RHS, an asymmetric link-prediction architecture in which the left-hand side (LHS) device tower leverages temporally valid watch-history message passing to capture collaborative signals, while the right-hand side (RHS) content tower is intentionally shallow with respect to the graph and encodes content solely from intrinsic features. The RHS tower does not use ID-based embeddings, content-side subgraphs, neighbor aggregation, or interaction-derived representations, forcing the content encoder to map intrinsic features into a collaborative-filtering-aware embedding space. After training, the learned content encoder generates embeddings for both warm and newly ingested content, enabling implicit graph completion through retrieval of warm surrogate neighbors. We further extend the same representation-completion principle to device cold-start by constructing cohort-based embeddings from demographic features. Large-scale online experiments demonstrate consistent relative improvements in content cold-start engagement, promotion speed, impression acquisition, and device cold-start engagement.
\end{abstract}



\maketitle

\section{Introduction}

Modern recommendation systems rely heavily on collaborative filtering (CF) signals extracted from user-content interactions. Graph-based models are particularly effective in this setting because they propagate behavioral information over interaction graphs and learn representations that capture high-order audience preference patterns. The cold-start problem arises because collaborative structure is missing for newly added nodes. In a streaming platform, this issue is acute: new titles must be surfaced before meaningful watch history accumulates, and newly activated devices must receive recommendations before personalized behavior is observed. This work is developed and deployed within Tubi, a large-scale ad-supported streaming platform serving hundreds of millions of users and a large catalog of content items, using the Kumo GNN platform as the graph modeling backbone.

The production setting imposes additional constraints beyond standard cold-start recommendation. The serving system requires both content and device embeddings that can be indexed, retrieved, and compared efficiently using approximate nearest-neighbor search. As a result, a solution cannot rely solely on pairwise device–content similarity scoring, nor can it construct embeddings for cold content through interaction-derived subgraphs on the target side. Newly ingested content has no watch-history edges, so the model must compute its embedding directly from intrinsic features, while ensuring that the resulting representation remains aligned with the collaborative structure of the embedding space.

We formulate this task as a representation-completion problem on a temporal bipartite graph of devices and content. For content cold-start, the goal is to learn an inductive function that maps metadata, taxonomy, and LLM-based semantic features into a CF-aware content embedding, even when the content node has zero observed edges. For device recommendation, the model also learns a device representation from device features and temporally valid watch history. For device cold-start, where no such history exists, we approximate the missing representation using demographic and contextual cohort priors. In this view, cold-start recommendation is an extreme form of graph completion: the node exists and its side information is available, but the interaction structure around it is missing.


To address the content side of this problem, we propose Shallow-RHS, an asymmetric architecture for temporal link prediction. The left-hand side (LHS) encodes the querying device using device features and message passing over its historical watch context, allowing the model to learn collaborative structure from interaction-rich neighborhoods. The right-hand side (RHS) encodes target content using only intrinsic content features. This asymmetry prevents the target tower from memorizing warm titles and forces the content encoder to explain future device-content links from content features alone. As a result, the learned content encoder maps both warm and newly ingested zero-history titles into a shared CF-aware embedding space.
%
%
%
We further apply the same representation-completion principle to device cold-start by assigning newly activated devices to demographic cohorts and using the cohort-level average of learned warm-device embeddings for retrieval. Together, these mechanisms provide a practical framework for reconstructing missing representations on both sides of the device-content graph. 
The main contributions of this work are: 
\begin{itemize}
    \item We formulate content cold-start as an inductive graph-completion problem, where zero-edge content nodes require feature-based representation inference beyond observed graph structure.
    \item We propose Shallow-RHS, an asymmetric temporal link-prediction architecture that produces device and content embeddings while keeping the RHS content tower free of content-side subgraphs and interaction-derived representations.
    \item We introduce an implicit graph-completion procedure that embeds newly ingested content with the learned content encoder and retrieves warm surrogate neighbors in the CF-aware content space.
    \item We extend the same representation-completion principle to device cold-start using demographic cohort priors and CF-aware retrieval.
    \item We validate the approach through large-scale online experiments demonstrating gains in total view time, cold-content promotion speed, impression acquisition, and first-touch device engagement.
\end{itemize}

\section{Related Work}

\subsection{Cold-start Recommendation}

The cold-start problem arises when a recommender model needs to rank new users or new items with little or no historical interaction data. This setting is especially challenging for collaborative filtering methods, since matrix factorization and related latent-factor models infer user and item representations primarily from observed user-item co-occurrence. When an item has no interactions, its collaborative representation is either undefined or poorly estimated, which prevents the recommendation model from generating meaningful representations for it in the learned embedding space 
\citep{schein2002coldstart,koren2009matrixfactorization}.

A common class of solution methods uses content-based recommendation, where item metadata, text, audio, images, or other intrinsic features are used to represent new items~\citep{lops2011content,oord2013deepcontent}. These methods naturally support item cold-start because they do not require historical user-item interactions. However, purely content-based similarity often reflects semantic-level similarity rather than audience preference. This creates a semantic-collaborative gap: content features provide generalization, but do not directly encode the behavioral signals captured by collaborative filtering methods.

Hybrid methods combine collaborative filtering with semantic information. Early approaches include probabilistic content-CF models for cold-start recommendation~\citep{schein2002coldstart}, collective matrix factorization over multiple relations~\citep{singh2008cmf}, and collaborative topic regression, which combines topic models with collaborative filtering for recommending newly published articles~\citep{wang2011ctr}. While these approaches improve over pure content-based recommendation methods, many still assume either partial interaction history, support examples, auxiliary relations, or warm item embeddings that can be directly reconstructed. Our work focuses on the strict item cold-start setting for newly added content with zero observed user-content interactions. 

\subsection{GNN-based Recommendation Systems}

Item recommendation problem can be naturally represented as a graph learning task, where user (device) nodes and item (content) nodes form a bipartite interaction graph, where the recommendation task is formulated as a link prediction problem. This formulation offers several advantages as the multi-hop graph structure naturally captures collaborative filtering signals. Furthermore, Graph Neural Networks (GNNs) are well suited to this problem setting where message passing aggregates information from local and higher-order neighborhoods while incorporating node and edge features. Popular GNN methods include PinSage which scales graph convolutional networks to web-scale graphs by combining random-walk-based neighborhood construction with localized graph convolutions~\citep{pinsage2018}. NGCF explicitly propagates user and item embeddings over the user-item interaction graph to encode high-order collaborative signals~\citep{ngcf2019}. ContextGNN frames recommendation task as a link prediction problem and combines local graph context with two-tower retrieval to improve ranking beyond pair-agnostic embeddings~\citep{contextgnn2024}.

\section{Methodology}

\subsection{Problem Formulation}

We formulate the cold-start content recommendation problem as a temporal link prediction task on a bipartite interaction graph. 
%
%
Let $\mathcal{G}_{\leq t}= \left(\mathcal{D} \cup \mathcal{C}, \mathcal{E}_{\leq t}, \mathbf{X}_{\mathcal{D}},
    \mathbf{X}_{\mathcal{C}}\right)$ 
%
%
%
denote the temporal graph observed up to timestamp $t$. The node set is bipartite: $\mathcal{D}$ denotes devices, or equivalently user identities in the streaming platform, and $\mathcal{C}$ denotes content nodes, also referred to as programs or titles. Edges only occur between device and content nodes. Each edge is a timestamped watch event defined as follows:
\begin{equation}
    e = (d,c,\tau,\mathrm{TVT}_{d,c,\tau}),
    \qquad
    d \in \mathcal{D},\; c \in \mathcal{C},\; \tau \leq t,
\end{equation}
where $\mathrm{TVT}_{d,c,\tau}$ is an edge feature corresponding to the total viewing time of device $d$ on content $c$ at timestamp $\tau$. Thus, the edge set $\mathcal{E}_{\leq t}$ is defined as:
\begin{equation}
    \mathcal{E}_{\leq t}
    \subseteq
    \mathcal{D} \times \mathcal{C} \times \mathbb{R}_{+} \times \mathbb{R}_{+}.
\end{equation}
We distinguish between warm and cold content nodes based on the graph connectivity at timestamp $t$. Warm content nodes are content nodes with observed watch edges: 
\begin{equation}
    \mathcal{C}_{warm}(t)
    =
    \{c \in \mathcal{C}
    \mid
    \exists d,\tau \leq t
    \text{ such that } (d,c,\tau,\cdot) \in \mathcal{E}
    \},
\end{equation}
On the other hand, cold content nodes correspond to content nodes with no observed watch event edges in the graph:
\begin{equation}
    \mathcal{C}_{cold}
    =
    \{c
    \mid
    (d,c,\tau,\cdot) \notin \mathcal{E},
    \forall d,\tau \}.
\end{equation}
As a result, such nodes cannot receive collaborative signals through standard graph message passing, and their representations are undefined under conventional graph-based recommendation models.

Each node in the graph is associated with heterogeneous features. A device feature vector $\mathbf{x}_d \in \mathbf{X}_{\mathcal{D}}$ may include device type, country, platform join time, and available demographic or account-level attributes such as age and gender. A content feature vector $\mathbf{x}_c \in \mathbf{X}_{\mathcal{C}}$ may include title text, content type, language, production year, duration, genre, structured metadata, and dense semantic embeddings derived from metadata, synopsis, or program scripts using pre-trained large language models. 

For a device $d$, we define its historical context up to timestamp $t$ as follows:
%
\begin{equation}
    \mathcal{N}^{<t}(d)
    =
    \{(c,\tau,\mathrm{TVT}_{d,c,\tau})
    \mid
    (d,c,\tau,\mathrm{TVT}_{d,c,\tau}) \in \mathcal{E},\; \tau < t
    \}.
\end{equation}
By extending the historical context $\mathcal{N}^{<t}(d)$ to include the multi-hop neighborhood of device $d$, we capture the viewing patterns from other users with overlapping watch histories, i.e. which programs did other users with similar watch histories also watch. 
The central learning problem is to infer a collaborative-filtering-consistent representation for zero-edge content nodes using only their intrinsic features. To do so, we need to learn two embedding functions defined as follows:
\begin{equation}
    \mathbf{z}_d(t)=f_{\theta_d}(\mathbf{x}_d, \mathcal{N}^{<t}(d))
    \qquad
    \mathbf{z}_c  = f_{\theta_c}(\mathbf{x}_c)
\end{equation}
The device embedding $\mathbf{z}_d(t)$ 
depends on the device’s historical watch subgraph \(\mathcal{N}^{<t}(d)\) as well as the device input features. In contrast, the content embedding $\mathbf{z}_c$ is constrained to depend only on intrinsic content features. This asymmetric constraint is critical as the content function \(f_{\theta_c}\) must be applicable to both warm and cold content, including newly ingested titles with no historical watch edges.

We therefore formulate the cold-start recommendation task as an extreme graph completion problem: the cold content node exists, but the interaction subgraph around it is absent. Given a device's historical context, the task is to rank the content items that the device will watch during a future window $(t,t+\Delta]$. The set of positive labels for this task is defined as:
%
\begin{equation}
    \mathcal{P}_{d}^{t,\Delta}
    =
    \{c \in \mathcal{C}_{warm}(t)
    \mid
    \exists \tau \in (t,t+\Delta]
    \land (d,c,\tau,\cdot) \in \mathcal{E}
    \},
\end{equation}
where $\mathcal{C}_{warm}(t)$ is the set of warm content nodes available at timestamp $t$. 
The temporal link-prediction score is defined as follows:
\begin{equation}
    s(d,c,t)
    =
    \cos(\mathbf{z}_d(t), \mathbf{z}_c)
    =
    \frac{\mathbf{z}_d(t)^{\top}\mathbf{z}_c}
    {\|\mathbf{z}_d(t)\|_2\|\mathbf{z}_c\|_2}.
\end{equation}

Although supervision labels are derived from interaction edges between devices and warm content nodes, the design of $f_{\theta_c}$ forces it to learn to explain collaborative behavior from content features alone. As a result, the learned content encoder acts as a semantic-to-collaborative bridge: it maps zero-history items into an embedding space shaped by historical user behavior. This formulation turns cold-start recommendation into an inductive graph-completion problem, where rather than requiring observed edges for every new item, we learn a feature-based embedding function into a collaborative space, allowing cold content to be encoded immediately upon being incorporated into the graph.

\subsection{Shallow-RHS Architecture}

In this section, we introduce our proposed Shallow-RHS architecture, which is specifically designed to tackle the item cold-start task formalized above, where newly added content nodes have intrinsic feature vectors ($\mathbf{x}_c$) available but no watch-history edges. The proposed Shallow-RHS also satisfies the requirement of our serving system that an embedding $\mathbf{z}_c$ can be retrieved immediately and used to promote the corresponding new content on the platform. The Shallow-RHS model therefore must learn not only to predict temporal device-content links, but also to output representative embedding functions for devices and content, with the content-side function remaining valid for zero-history titles.


In the proposed Shallow-RHS, we adopt an asymmetric two-tower architecture for temporal link prediction. 
The left-hand side (LHS) encodes the querying device using device features and temporally valid watch-history message passing over $\mathcal{N}^{<t}(d)$, allowing it to capture collaborative filtering signals from interaction-rich neighborhoods. In contrast, the right-hand side (RHS) is intentionally shallow with respect to the graph: it encodes target content using only intrinsic content features, without access to interaction-derived information. 
As a result, the learned content encoder is an inductive mapping from intrinsic content features to a collaborative-filtering-aware embedding space.

Both towers encode raw heterogeneous features with a HeteroTF-style tabular encoder built using PyTorch Frame's FTTransformer \citep{pytorchframe2024}. Semantic-type-specific encoders map numerical, categorical, timestamp, text, and precomputed embedding columns into shared column embeddings; FT-Transformer self-attention then mixes information across columns before pooling the result into a node representation. 

The RHS content representation is shallow with respect to the graph and is defined as follows:
\begin{equation}
    \mathbf{z}_c
    =
    f_{\theta_c}(\mathbf{x}_c)
    =
    \mathrm{HeteroTF}_{C}(\mathbf{x}_c)
\end{equation}
No message passing is applied from devices into content nodes on the RHS, and the content embedding used for scoring is not computed from the users or devices connected to that content. This 
is central to the cold-start generalization requirement where the same function $f_{\theta_c}$ can be applied to warm and cold content as it depends solely on intrinsic content features.

On the other hand, the LHS device representation combines device features with temporally valid historical watch events. At first, we compute an initial embedding of the device node intrinsic features as follows: 
\begin{equation}
    \mathbf{h}_{d}^{(0)}
    =
    \mathrm{HeteroTF}_{D}(\mathbf{x}_d).
\end{equation}

Then, for each historical edge $(d,c,\tau,\mathrm{TVT}_{d,c,\tau})$ with $\tau < t$, the model constructs a content-to-device message using the content encoder output and edge features:
\begin{equation}
    \mathbf{m}_{d \leftarrow c,\tau}^{(\ell)}
    =
    \psi^{(\ell)}
    \left(
    \mathbf{z}_c,\;
    \gamma(\mathrm{TVT}_{d,c,\tau}),\;
    \eta(t-\tau)
    \right),
\end{equation}
where $\gamma(\cdot)$ encodes total viewing time and $\eta(\cdot)$ encodes temporal recency. Given a message multiset $\mathcal{M}_{d}^{<t}$, we use an aggregation block: 
%
%
%
\begin{equation}
    \mathrm{Aggr}(\mathcal{M}_{d}^{<t})
    =
    \mathrm{MLP}
    \left(
    \mathrm{Concat}
    \left(
    \mathrm{Aggr}_{1}(\mathcal{M}_{d}^{<t}),
    \ldots,
    \mathrm{Aggr}_{R}(\mathcal{M}_{d}^{<t})
    \right)
    \right)
\end{equation}
where each $\mathrm{Aggr}_{r}(\cdot)$ denotes a permutation-invariant aggregator over the message multiset, such as mean or max operators.
The device representation is then updated by one or more heterogeneous aggregation layers:
\begin{equation}
    \mathbf{h}_{d}^{(\ell+1)}
    =
    \mathrm{MLP}^{(\ell)}
    \left(
    \mathrm{Concat}
    \left(
    \mathbf{h}_{d}^{(\ell)},
    \mathrm{Aggr}(\mathcal{M}_{d}^{<t})
    \right)
    \right).
\end{equation}
Finally, the output of the last aggregation layer $L$ represents the final device embedding $\mathbf{z}_d(t) = \mathbf{h}_{d}^{(L)}$.
%

The training process utilizes temporal softmax link prediction loss where for each device-timestamp pair $(d,t)$, the positives correspond to future watched content in $\mathcal{P}_{d}^{t,\Delta}$. On the other hand, the negatives $\mathcal{N}_{d}^{t,\Delta}$ are candidate content nodes not watched by $d$ in the prediction window. In particular, the loss function is defined as follows: 
%
\begin{equation}
    \mathcal{L}
    =
    -
    \sum_{(d,t)}
    \sum_{c^{+} \in \mathcal{P}_{d}^{t,\Delta}}
    \log
    \frac{
    \exp(s(d,c^{+},t)/\tau_s)
    }{
    \exp(s(d,c^{+},t)/\tau_s)
    +
    \sum_{c^{-} \in \mathcal{N}_{d}^{t,\Delta}}
    \exp(s(d,c^{-},t)/\tau_s)
    },
\end{equation}
where $\tau_s$ is a temperature hyper-parameter. The inputs to the device tower are restricted to edges with timestamps $\tau < t$, while labels are drawn only from $(t,t+\Delta]$. This prevents label leakage 
during 
the message-passing process. 
The entire model is trained end-to-end to minimize loss $\mathcal{L}$. 

\subsection{Implicit Graph Completion via Surrogate Neighbors}

After the training of the Shallow-RHS model is complete, the learned content encoder is used to generate embeddings for all warm or newly added cold content nodes as follows: 
\begin{equation}
    \mathbf{z}_c = f_{\theta_c}(\mathbf{x}_c).
\end{equation}
%
%
Subsequently, we build an approximate nearest-neighbor index over warm content embeddings $\mathcal{Z}_{warm} = \{\mathbf{z}_{w} \mid w \in \mathcal{C}_{warm}\}.$
%
%
Then, for each cold content node $c \in \mathcal{C}_{cold}$, we retrieve its top-$M$ warm surrogate neighbors as follows:
\begin{equation}
    \mathcal{S}_{M}(c)
    =
    \operatorname{TopM}_{w \in \mathcal{C}_{warm}}
    \cos(\mathbf{z}_c,\mathbf{z}_{w}).
\end{equation}

This 
implicitly completes the graph around cold content nodes. Rather than adding synthetic device-content edges, this approach connects cold content 
to nearby warm content in the learned embedding space. The surrogate neighbor set $\mathcal{S}_{M}(c)$ provides an interpretable bridge from cold content to the existing interaction-rich warm content catalog, allowing cold titles to be promoted 
alongside warm titles. 

In production, these surrogate neighbors are not treated as ground-truth watch edges. Instead, they are used as serving-time behavioral proxies: a cold title inherits evidence from nearby warm titles in the CF-aware space for promotion, retrieval, or calibration decisions. Thus, surrogate completion operationalizes the missing neighborhood of a cold item without modifying the observed interaction graph.


\subsection{Cold-Start Device Representation via Demographic Priors}

The previous sections address the content side of the cold-start problem by generating 
item representations from semantic priors. The same principle can be applied to the query (device) side of the bipartite graph. New devices also correspond to nodes with missing graph structure: they have no watch-history neighborhood and, therefore, cannot be embedded 
through the LHS graph encoder. Instead of semantic priors, device cold-start relies on demographic and contextual priors.

We cluster 
warm devices into approximately $K_D$ demographic cohorts based on available device and account-level attributes. For each cohort $g$, we compute a representative 
embedding by averaging the learned device embeddings of warm devices assigned to that cohort:
\begin{equation}
    \mathbf{z}_g =
    \frac{1}{|\mathcal{D}_g|}
    \sum_{d \in \mathcal{D}_g}
    \mathbf{z}_d .
\end{equation}
For each newly activated device $d_{new}$, we infer its cohort assignment $g(d_{new})$ from demographic features and use $\mathbf{z}_{g(d_{new})}$ as its initial device representation:
\begin{equation}
    \mathbf{z}_{d_{new}} = \mathbf{z}_{g(d_{new})}.
\end{equation}

This enables immediate retrieval by performing an approximate nearest-neighbor search between the cohort-based device embedding and the CF-aware content embeddings learned by the Shallow-RHS model. Thus, item cold-start and device cold-start share the same underlying methodology: missing graph representations are reconstructed from side-information priors and then used in the collaborative embedding space.

\section{Experimental Results}

\subsection{Graph Construction from Relational Data}

The proposed Shallow-RHS architecture is implemented within the Kumo GNN platform and adapted for Tubi’s production cold-start recommendation setting. We construct the temporal bipartite graph using one year of watch history logs from Tubi's viewership.
Let $t_0$ denote the graph construction cutoff time, $T=365$ days denote the historical lookback window, and 
$r_{\min}$ denote the minimum total viewing time threshold. We retain only watch events with sufficient viewing duration:
\begin{equation}
    \mathcal{E}_{\mathrm{train}}
    =
    \left\{
    (d,c,\tau,r) \in \mathcal{E}
    \mid
    t_0 - T \leq \tau \leq t_0,\;
    r \geq r_{\min}
    \right\},
\end{equation}
where $d \in \mathcal{D}$ is a device node, $c \in \mathcal{C}$ is a content node, $\tau$ is the watch timestamp, and $r=\mathrm{TVT}_{d,c,\tau}$ is the total viewing time. Filtering short views removes weak or accidental interactions and discards noisy edges in the graph. We construct the temporal bipartite graph from production-scale watch-history logs collected from Tubi’s recommendation system. The resulting graph contains hundreds of millions of 
device nodes, 
hundreds of thousands of content nodes, and 
billions of temporal watch edges. Multiple watch events between the same device and content nodes are preserved as temporal edges rather than collapsed into a single static edge, allowing the model to condition on recency and repeated engagement.

We separate historical context construction from supervision signal. The full one-year graph is used to construct temporal device watch histories, but link-prediction labels are sampled only from the last K-many days 
before $t_0$ defined as $\mathcal{T}_{\mathrm{sup}}=  [t_0 - 
K\text{ days}, t_0].$ 
For each training anchor time $t \in \mathcal{T}_{\mathrm{sup}}$, the device tower observes only historical events with $\tau < t$, while positives are future watch events in $(t,t+\Delta]$. This 
enables the model to use long-range behavioral context while emphasizing recent catalog trends and recent device preferences. 

Device features capture geography, platform, and available account-level attributes. In particular, $\mathbf{x}_d$ may include country or region, device type and application platform. Content features capture intrinsic and external title information, including title text, content type, genre taxonomy, language, release or production year, duration, maturity rating, cast and director metadata, popularity or quality signals such as external ratings and vote counts, and LLM-based semantic embeddings derived from program scripts or metadata. 

Feature coverage is a critical empirical constraint. In the initial system, pre-trained LLM
-based embeddings covered only a small portion of 
content nodes, and several external metadata fields such as ratings, vote counts, budget, and box-office signals had partial coverage. We treat this limited coverage as a representation quality issue: if critical content features are absent, the learned mapping $f_{\theta_c}(\mathbf{x}_c)$ has less information with which to align semantic content representations to collaborative behavior. This observation motivates the progressive feature-enrichment experiments described in the following sections.

Given the size of the temporal graph, to ensure training scalability, 
we sample a bounded temporal device history for each training instance retaining only the most recent $K$ 
content interactions. 
The initial September experiment used $D$
-dimensional output embeddings. After content-side feature enrichment in the October experiment, we doubled 
the output embedding dimension 
to improve representation capacity for richer program metadata and script-derived semantic features.

\subsection{Evaluation Setup}

All experiments are conducted as randomized online A/B tests within Tubi’s production recommendation system. The primary objective is to measure improvements in user engagement and cold-start content exposure under real serving conditions. 
We report the following key engagement metrics:
\begin{itemize}
    \item \textbf{Total View Time (TVT):} The average daily viewing time per device, which reflects overall user engagement with recommended content.
    \item \textbf{Qualified View Days:} The number of days on which a device achieves a minimum engagement threshold (e.g., at least K 
    minutes of viewing), capturing sustained user activity.
    \item \textbf{Conversion Metrics:} Short-term engagement signals such as homepage 5-minute conversion, measuring whether users quickly engage with recommended content.
\end{itemize}

%
In each A/B test, treatment variants are compared against a production baseline. The metrics are aggregated over sufficiently long evaluation windows to ensure statistical stability. The evaluation focuses on both global performance and cold-start-specific behavior, including promotion speed and early exposure of newly ingested content. This evaluation setup allows us to measure not only the quality of offline embedding but also the real-world impact on recommendation performance under cold-start conditions.

\subsection{Progressive Bootstrapping and Ablation Analysis of Content Cold-Start}

We evaluated the Shallow-RHS model through a sequence of online A/B experiments designed to progressively transfer collaborative-filtering signal into the content embedding space. Over multiple phases, the central mechanism is: improving content-side feature coverage and quality improves the learned content embedding which in turn improves the alignment between intrinsic content semantics and graph-derived collaborative behavior.


\subsubsection{Phase 1: Semantic-to-CF Replacement}
In the first experiment, we replaced the prior content-only semantic-based 
representation (Control) with Shallow-RHS embeddings (Treatment). The Shallow-RHS model used content metadata, pre-trained LLM
-based summary 
embeddings where available, and graph-derived device histories during training. Unlike raw semantic embeddings, the resulting content embeddings were optimized through temporal device-content link prediction and therefore reflected collaborative preference structure. This phase achieved a $+0.10\%$ global TVT lift and increased cold-title promotion speed by $13\%$. 
Given Tubi’s viewership scale, a lift of this magnitude represents a major win, especially considering the historical difficulty of driving engagement gains through cold-start content optimizations.
These results indicate that the CF-aware embedding space learned by Shallow-RHS is more effective for cold-start promotion than a purely semantic embedding space.

\subsubsection{Phase 2: Metadata Coverage and Decision Calibration}
In the second experiment, we improved both the input feature coverage and the promotion decision logic while retaining the same Shallow-RHS architecture. The content schema was enriched with higher-coverage genre features, budget metadata, box-office and popularity signals, and improved pre-trained LLM
-based embedding coverage. We also replaced the previous coarse tier-score decision rule with a calibrated binary promotion-eligibility signal used by the serving system. 
This phase produced a $+0.16\%$ global TVT lift, suggesting that better metadata coverage increased the effectiveness of the semantic-to-CF alignment.

\begin{table*}[t]
\centering
\begin{tabular}{lcccccc}
\toprule
Model & GNN & Semantic & Calibration & Feature Enhancement & Surrogate & TVT Lift \\
\midrule
V1 & $\checkmark$ & $\checkmark$ & $\times$ & $\times$ & $\times$ & +0.1\% \\
V2 & $\checkmark$ & $\checkmark$ & $\checkmark$ & $\times$ & $\times$ & +0.16\% \\
V3 & $\checkmark$ & $\checkmark$ & $\checkmark$ & $\checkmark$ & $\times$ & +0.42\% \\
V4 & $\checkmark$ & $\checkmark$ & $\checkmark$ & $\checkmark$ & $\checkmark$ & +0.17\% \\
\bottomrule
\end{tabular}
\caption{Relative TVT lift across progressive system iterations (reported as percentage lift over baseline)}
\label{tab:ablation}
\end{table*}

\subsubsection{Phase 3: Content-Side Feature Completion}

In the third experiment, we further expanded content feature coverage and reduced missing values across key content attributes. This phase achieved approximately 10\% higher feature coverage across the catalog, with near-complete coverage for the most important semantic features. The resulting similarity distributions were smoother and better calibrated for cold-start retrieval, indicating that the learned content space was less affected by missing feature artifacts. New titles reached key exposure milestones more quickly than the control group, and the experiment produced a $+0.42\%$ global TVT lift. This phase delivered the most pronounced improvement among the evaluated rollout stages, highlighting the importance of content-side representation quality for zero-edge content generalization.

\subsubsection{Phase 4: Surrogate Completion and Semantic Enrichment}
In the fourth experiment, we further enriched and expanded the content representation by introducing target-audience and demographic descriptors, initialized with pre-trained OpenAI-based LLM embeddings. This provided richer contextual information, leading to more robust training for the Shadow-RHS model. Furthermore, to address the few long-tail titles where our primary strategy fell short, we utilized these pre-trained LLM embeddings to compute and fulfill their surrogate representations. This acted as a fallback mechanism to ensure comprehensive surrogate completion across all titles. Consequently, this joint strategy delivered a $+0.17\%$ lift in global TVT

Overall, these experiments demonstrate that content cold-start performance depends jointly on the feature coverage and semantic quality of content-side features. The repeated performance gains support the hypothesis that cold-start recommendation can be treated as an implicit graph completion task: the model first learns a CF-aware semantic embedding function from warm temporal interactions, then applies that function to newly ingested content with no observed device edges. Table~\ref{tab:ablation} summarizes the incremental contribution of each major component introduced during the phased 
rollout. While the four variants are evaluated as successive online system iterations rather than isolated offline component removals, they provide a practical ablation of the cold-start pipeline under real serving conditions.



\subsection{Similarity Distribution Analysis}

To better understand how the proposed architecture reshapes the embedding space for cold-start content, we analyze the distribution of pairwise similarity scores under different embedding strategies. The reported similarity scores are normalized embedding similarities computed over aggregate cold-start evaluation cohorts.

In traditional collaborative filtering (CF) embeddings, cold-start items lack interaction history and therefore fail to establish meaningful relationships with existing content. As a result, similarity scores involving cold items tend to concentrate in a low-similarity regime, effectively limiting their visibility in retrieval and ranking stages.

In contrast, purely semantic embeddings (e.g., pre-trained LLM-based representations derived from metadata and scripts) produce a more centralized similarity distribution. While this provides generalization and reasonable coverage, the scores remain relatively neutral, reflecting semantic closeness rather than true audience preference. Consequently, such embeddings lack the behavioral signals necessary for effective recommendation.

\begin{figure}[h]
    \centering
    \includegraphics[width=0.5\textwidth]{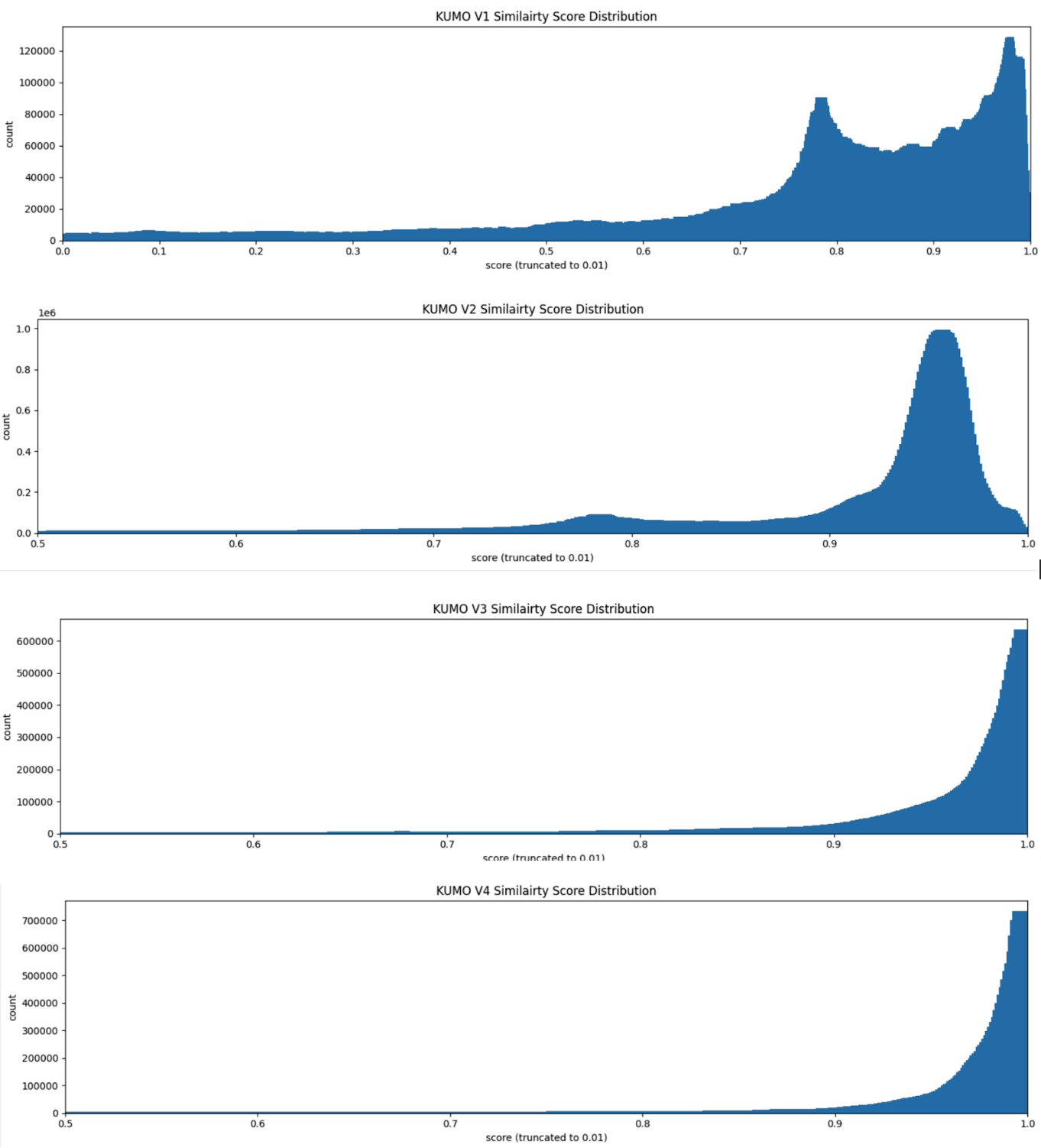}
    \caption{Similarity Score Distribution Progress}
    \label{fig:scoredist}
\end{figure}

The Shallow-RHS model produces a markedly different distribution. As shown in Figure~\ref{fig:scoredist}, similarity scores shift toward a higher-similarity regime, resulting in a right-skewed distribution. This shift indicates that the learned content embeddings no longer represent purely semantic similarity, but instead capture collaborative filtering structure induced by historical user interactions.

This distributional transformation provides direct evidence for our central hypothesis: the proposed architecture successfully aligns semantic representations with collaborative filtering behavior. As a result, cold-start content can be embedded into a CF-aware space, enabling it to connect meaningfully to the existing content catalog and be surfaced effectively in recommendation.

\subsection{Device Cold-Start Evaluation}

We also evaluated the proposed model on newly activated devices with little or no watch history. For these devices, the model used demographic cohort embeddings to retrieve candidate content from the CF-aware content embedding index. This allowed the retrieval layer to provide personalized content candidates before sufficient device-specific interactions are observed. In an online A/B test, this device cold-start retrieval strategy improved multiple engagement metrics. Qualified view days increased by $+0.29\%$, capped daily total view time increased by $+0.39\%$, and homepage five-minute conversion increased by $+0.43\%$. These gains highlight that reconstructing missing device representations from demographic priors can improve first-touch recommendation quality, complementing the item-side cold-start improvements from semantic-to-CF alignment.

\section{Conclusion}

%

We presented a production-scale approach to cold-start recommendation that treats cold-start as missing graph structure. 
In a temporal bipartite device-content graph, newly added content and newly activated devices lack the interaction neighborhoods that collaborative filtering models normally use to construct representations. The central problem is therefore representation completion: producing useful embeddings for zero-edge nodes while keeping those embeddings compatible with the collaborative space used by the retrieval system.
For content cold-start, Shallow-RHS transfers collaborative signal from temporal device-content interactions into an inductive content encoder. The LHS device tower uses graph-based historical context, while the RHS content tower remains feature-only and does not rely on a content-side subgraph 
or interaction-derived features. This asymmetric constraint allows the same content encoder to generate CF-aware embeddings for both warm and newly ingested zero-history content. Surrogate-neighbor retrieval then operationalizes implicit graph completion by connecting cold content to nearby warm content in the learned embedding space. For device cold-start, demographic cohort embeddings extend the same principle to the query side of the graph. 
Online experiments show that the proposed framework improves both content and device cold-start performance, including global TVT gains, faster cold-content exposure, and improved first-touch device engagement.

\bibliographystyle{ACM-Reference-Format}
\bibliography{acmart}

\appendix

\section{Architecture Diagram}

\begin{figure*}[p]
    \centering
    \includegraphics[width=\textwidth,height=0.99\textheight,keepaspectratio]{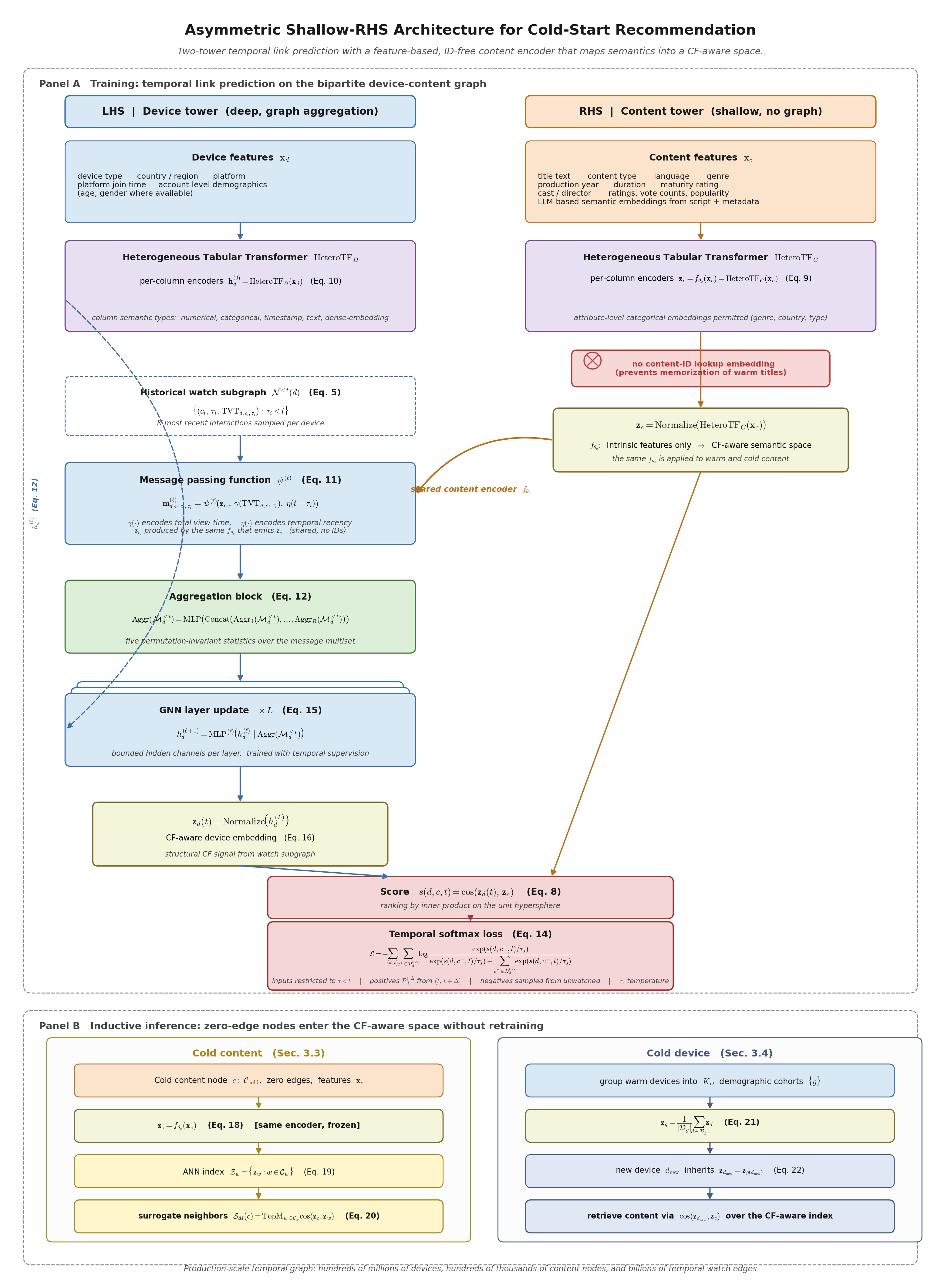}
    \caption{Overview of the Shallow-RHS architecture for content and device cold-start recommendation.}
    \label{fig:appendix_architecture}
\end{figure*}


\end{document}